\documentclass[a4paper,11pt]{article}
\usepackage{graphicx}
\usepackage{url}

\title
{
An Introduction to Logical Operations on Classical and Quantum Bits
\thanks{The authors are undergraduate students of Computer Science at 
the Catholic University of Petr\'opolis. They are also members of Grupo de
F\'{\i}sica Te\'orica Jos\'e Leite Lopes.}
}
\author
{
F.L.~Marquezino and R.R.~Mello~J\'unior\\
\\
CBPF - Centro Brasileiro de Pesquisas F\'{\i}sicas \\
CCP - Coordena\c{c}\~ao de Campos e Part\'{\i}culas \\
Av. Dr. Xavier Sigaud, 150\\
22.290-180 Rio de Janeiro (RJ) Brazil\\
(CNPq Fellows/PIBIC)\\
{\it franklin@serraon.com.br, rui.rodrigues@inf.ucp.br}
}

\date{April, 2004}

\begin{document} 

\maketitle

\begin{abstract}
This article is a short review on the concept of information. We show the strong relation between Information Theory and Physics, and the differences between classical and quantum information, with emphasis in their manipulation through logical gates.%
This paper is intended to be read by non-specialists and undergraduate students of Computer Science, Mathematics and Physics, with knowledge in Linear Algebra and Quantum Mechanics.
\end{abstract}

\section{Introduction}

In a previous paper~\cite{consid}, we expose the physical character of information, making it clear that every information must be represented by a physical system. However, another question that may arise is how to interfere on the information being represented? This subject will be studied in this article, under the viewpoints of both Classical and Quantum Computation.

Sections \ref{sec:classical} and \ref{sec:quantum} give an introduction to two different models of computation. The former is the traditional circuit model, compatible with the idea of Turing machines, but more adequate to the purposes of this paper. The latter is the corresponding concept that uses properties of Quantum Mechanics, instead of Classical Mechanics. 

As soon as we learn how to perform simple operations with a certain physical system, it is very important to learn how to keep these operations reliable. In Section~\ref{sec:EC}, we briefly introduce classical and quantum error detection and correction.

Here, we assume that the reader already has a certain backgroung on Quantum Information. If necessary, we recommend the reading of~\cite{consid} or~\cite{Portugal03} before starting reading Section~\ref{sec:quantum}.

\section{On classical circuits}\label{sec:classical}

The Turing machine is an important concept, developed by the british mathematician Alan Turing~\cite{Turing}. Any calculation using a classical physical system, can be carried out by a Turing machine, according to the Church-Turing thesis~\cite{Shapiro}:

\begin{quote}
``Any process that is effective or algorithmic in nature defines a mathematical function belonging to a specific well-defined class, known variously as the recursive, the $\lambda$-definable, or the Turing computable functions.''
\end{quote}

Here, we use a different, yet equivalent formalism, which is the circuit of gates and wires. The proof for the equivalence between Turing machines and circuits can be achieved through the notion of \textit{uniform circuit family}. The reader may found this proof in many books on Computer Science or Quantum Computation, including~\cite{Chuang00}.

We shall begin our study of the classical circuits by introducing the notion of Boolean functions. A function $f:\{0,1\}^n \rightarrow \{0,1\}^m$ is called a Boolean function, due to the english mathematician George Boole.

Interesting information on this subject can be found in ~\cite{Maryland}, specially in the section on Combinational Logics. Another good reference is the book by Andrew Tanenbaum~\cite{Tanenbaum}.

We can define a Boolean function $f(x)$ which is equal to $0$ if $x=1$ and is equal to $1$ if $x=0$. This function corresponds to the logical operation \textbf{NOT}. We can write the result of $f(x)$ as $\neg x$ or $\bar{x}$.

Another interesting function should be $f(x,y)$, which is equal to $1$ if $x=y=1$, and $0$ otherwise. This is the logical gate \textbf{AND}. Similarly, we could define a function $f(x,y)$, equal to $0$ if $x=y=0$ and $1$ otherwise, which corresponds to the logical gate \textbf{OR}. We can also write an AND operation as a product and an OR operation as a sum. For instance, $f(x,y)=xy$, meaning ``$x$ and $y$'', or $f(x,y)=x+y$, meaning ``$x$ or $y$''.

These functions can be represented by a table, called the truth table. See Tables \ref{tab:not}, \ref{tab:and} and \ref{tab:or}, for the functions NOT, AND and OR, respectively. We can build any circuit with these three logical operations.

\vspace{3.0mm}
\begin{table}[h]
\centering
\footnotesize
\begin{tabular}{|c|c|}
\hline
 $x$ & $\neg x$ \\
\hline
0 & 1 \\
\hline
1 & 0 \\
\hline
\end{tabular}
\caption{Logical operation NOT.}\label{tab:not}
\end{table}

\vspace{3.0mm}
\begin{table}[h]
\centering
\footnotesize
\begin{tabular}{|c|c|c|}
\hline
 $x$ & $y$ & $xy$ \\
\hline
0 & 0 & 0 \\
\hline
0 & 1 & 0\\
\hline
1 & 0 & 0\\
\hline
1 & 1 & 1\\
\hline
\end{tabular}
\caption{Logical operation AND.}\label{tab:and}
\end{table}

\vspace{3.0mm}
\begin{table}[h]
\centering
\footnotesize
\begin{tabular}{|c|c|c|}
\hline
 $x$ & $y$ & $x+y$ \\
\hline
0 & 0 & 0 \\
\hline
0 & 1 & 1\\
\hline
1 & 0 & 1\\
\hline
1 & 1 & 1\\
\hline
\end{tabular}
\caption{Logical operation OR.}\label{tab:or}
\end{table}

\begin{figure}
\centering
\includegraphics[width=210pt]{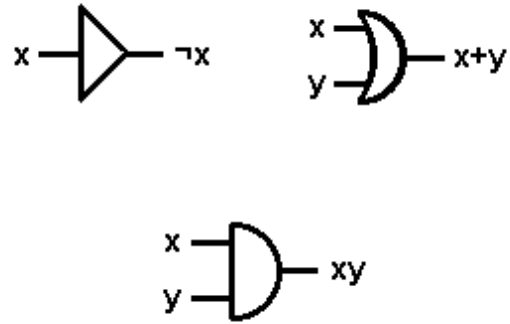}
\caption{Graphical representation for the logical operations NOT, OR and AND.}
\label{fig:gates}
\end{figure}

At this point we should finally introduce the circuit of gates and wires. In this model each gate has a graphical representation. Wires carry 0's and 1's, through space or time. See Fig.\ref{fig:gates}, for the functions NOT, OR and AND.

\begin{figure}
\centering
\includegraphics[width=190pt]{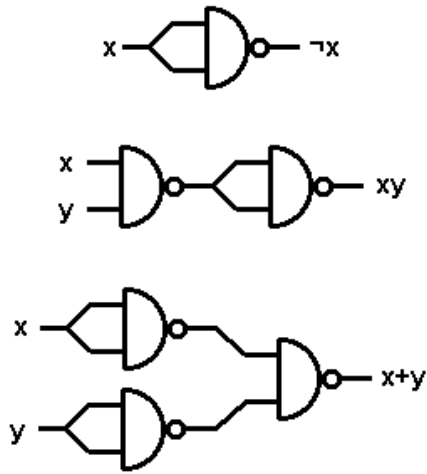}
\caption{We can use the NAND gate to build the gates AND, OR and NOT. Consequently, we can build any classical circuit using only NAND gates.}
\label{fig:nand-univ}
\end{figure}

The output of a gate can be the input of another gate\footnote{Normally, the output of a gate is not a input of the same gate, although it may be possible in some special cases, such as flip-flop circuits. Flip-flops 
are special circuits that work as memories, keeping the value of a bit even after the input signalization.}. So, let $f(x)$ be a Boolean function corresponding to the NOT logical operation, and let $g(x,y)$ be a Boolean function corresponding to the AND logical gate. We can define a function $h=f \circ g=\neg(xy)$. This function is very important in Computer Science, and is called NAND. Graphically, it can be represented by putting a small circle in the outcome of the AND gate, meaning that this result is negated (see Fig.~\ref{fig:nand-univ}). Any classical logical circuit can be designed using only the NAND logical gate, and the FANOUT\footnote{A ``gate'' that replicates the value of one bit in two different wires. It is often represented by a bifurcation in a wire.}. 

Another interesting logical gate is the NOR, defined as $f(x,y)=\neg (x+y)$, that is, an OR logical gate with its outcome inverted. As well as the NAND gate, this gate can also be used to build any logical circuit.

\begin{figure}
\centering
\includegraphics[width=190pt]{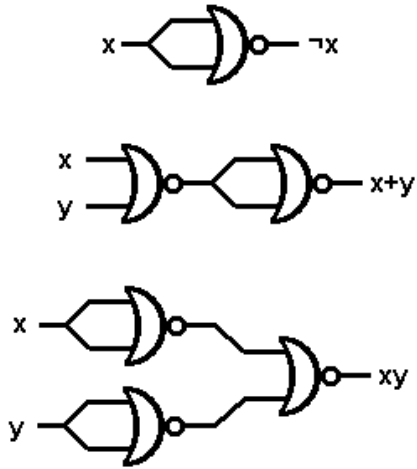}
\caption{We can also use the NOR gate to build the gates AND, OR and NOT. Consequently, the gate NOR is also universal.}
\label{fig:nor-univ}
\end{figure}

We cannot expect to create quantum circuits in a straightfoward way by simply using the same techniques of classical circuits. Firstly, because in classical circuits the universal gate is NAND, which is irreversible\footnote{If we know only the output of an irreversible gate, we do not necessarily know the inputs.}, while in quantum circuits all gates must be reversible. Not even FANOUT is possible in Quantum Mechanics, because of the non-cloning theorem~\cite{Wooters82}.

\section{On quantum circuits}\label{sec:quantum}

Similarly to Classical Computation, there must be a mathematical representation for the operations performed in the quantum physical system. Computer scientists are used to represent logical operations by Boolean functions. However, in Quantum Computation we will use a slightly different formalism, which will be more suitable. The reader must be familiar with Linear Algebra, because that is what we will use from now on.

A quantum state is represented by a vector in a Hilbert space (a complex vector space with inner product). Hence, if we consider an orthonormal basis composed by the states $|0\rangle$ and $|1\rangle$, the quantum state can be any of the form $\alpha |0\rangle + \beta |1\rangle$, where $\alpha$ and $\beta$ are complex numbers called amplitudes. Since this state must be a unit vector, we know that $\| \alpha \|^2 + \| \beta \|^2=1$. 

We can represent our computational basis by the state vectors

\begin{equation}
|0\rangle \equiv \left[\begin{array}{c}1\\0\end{array}\right] \mbox{ and } 
|1\rangle \equiv \left[\begin{array}{c}0\\1\end{array}\right],
\end{equation}
for example. This representarion may help the understanding of the calculations carried out in this section.

Operations on a quantum state must preserve its norm, so they are described by unitary matrices. In order to represent a logical operation on a qubit, one should find a unitary matrix that gives the desired result when applied to the qubit.

\begin{equation}
U | \psi \rangle = | \psi^{'} \rangle.
\end{equation}

To illustrate that, let us observe the equivalent of the NOT classical logic gate. We must find a unitary operator U, such that

\begin{eqnarray}
U | 0 \rangle = | 1 \rangle \\
U | 1 \rangle = | 0 \rangle,
\end{eqnarray}
and, since U is a linear operator,

\begin{equation}
U (\alpha | 0 \rangle + \beta | 1 \rangle) = \alpha U  | 0 \rangle + \beta U | 1 \rangle.
\end{equation}

The operator

\begin{equation}
\sigma_x = \left[
                   \begin{array}{cc}
                        0 & 1\\
                        1 & 0
                   \end{array}
               \right],
\end{equation}
one of the three Pauli matrices, corresponds to the unitary matrix that we are looking for.

Note that if we apply the operator $\sigma_x$ twice, we keep the original value of the qubit.

\begin{equation}
\sigma_x \sigma_x = \left[
                                     \begin{array}{cc}
                                          0 & 1\\
                                          1 & 0
                                   \end{array}
                          \right]
                          \left[
                                     \begin{array}{cc}
                                          0 & 1\\
                                          1 & 0
                                   \end{array}
                          \right]=
                          \left[
                                     \begin{array}{cc}
                                          1 & 0\\
                                          0 & 1
                                   \end{array}
                          \right]
\end{equation}
which is the identity matrix. 

Another interesting operator is the Hadamard gate:

\begin{equation}   
H= \frac{1}{\sqrt{2}}
   \left[
       \begin{array}{cc}
           1 & 1\\
           1 & 1
      \end{array}
  \right].\label{eq:hadamard}
\end{equation}

This gate is able to create superpositions in a qubit. Note that $H | 0 \rangle = \frac{1}{\sqrt{2}}(|0\rangle + |1\rangle)$ and $H | 1 \rangle = \frac{1}{\sqrt{2}}(|0\rangle - |1\rangle)$. 

Another interesting type of operator that we cannot forget mentioning is the controlled operator. When we have at least two qubits (a quantum register), we can define an operator that will interfere in a particular qubit if and only if some requirement is observed. For instance, let $\sigma_x^{c,t}$ be an operator that applies the NOT logic operation in the qubit $t$ if and only if the qubit $c$ is $|1\rangle$.

\begin{eqnarray}
\sigma_x^{c,t} (|0\rangle_c |0\rangle_t) = |0\rangle_c |0\rangle_t \\
\sigma_x^{c,t} (|0\rangle_c |1\rangle_t) = |0\rangle_c |1\rangle_t \\
\sigma_x^{c,t} (|1\rangle_c |0\rangle_t) = |1\rangle_c |1\rangle_t \\
\sigma_x^{c,t} (|1\rangle_c |1\rangle_t) = |1\rangle_c |0\rangle_t.
\end{eqnarray}

In this case, we say that qubit $t$ is the ``target qubit'' and qubit $c$ is the ``control qubit''. This gate is called CNOT and, similarly, it may be expressed as a unitary matrix.

\begin{equation}   
CNOT= 
   \left[
       \begin{array}{cccc}
           1 & 0 & 0 & 0\\
           0 & 1 & 0 & 0\\
           0 & 0 & 0 & 1\\
           0 & 0 & 1 & 0
      \end{array}
  \right].
\end{equation}

Similarly to Classical Computation, there are gates in Quantum Computation that can be used to build any possible circuit. It can be proven that an arbitrary unitary matrix $U$ can be decomposed into a product of two-level unitary matrices, i.e., matrices that act non-trivially on \textbf{no more than} two vector componets. One could also prove that single qubit gates and CNOTs can be used to achieve any two-level unitary matrix. Hence, it can be proven that single qubit gates and CNOT gates can implement an arbitrary operation and, therefore, they are universal in Quantum Computation.

The proof for the last paragraph is not straightfoward and would spend a reasonable space in this paper. If the reader is interested in studying this subject deeper, the main reference is~\cite{UnivGates}. 

\section{Error detection and correction: some brief remarks}\label{sec:EC}

An important area of Computer Science concerns the detection and correction of errors. If we intend to build a computer, no matter if a classical or a quantum-mechanical one, we have to deal with its errors, so that the calculations can be trusted.

In a classical digital computer, we can consider one single kind of error. Suppose we wish to send a bit through a channel. If there is noise in this channel, a bit $0$ can be changed to $1$, with probability $p$, and a bit $1$ can be changed to $0$, with probability $q$. So, the bit $0$ will keep correct with probability $1-p$, and the bit 1 with probability $1-q$. Normally, we consider $p=q$. In this case, both bits $0$ and $1$ have the same probability $p$ of being affected by the noise, and the same probability $1-p$ of being sent correctly. This channel is called a \textit{binary symmetric channel}.

Classical schemes for error detection and correction use some kind of redundancy. A very simplistic example would be repeating the same bit $n$ times. For instance

\begin{eqnarray}
0_L=000\\
1_L=111
\end{eqnarray}
with $n=3$. 

In this case, if we wish to represent the number 5 (101 in the binary system), we must repeat each logical bit three times:

\begin{equation}
(101)_L=111000111.
\end{equation}

If some noise inverts a single bit in a triplet, we can easily repair the information. For instance, if the above number is changed to $111010110$, we know that the triplets $010$ and $110$ are wrong.  We can recover the information by performing the \textit{majority vote}. So, the triplet $010$ was originally $000$ and the triplet $110$ was $111$. One could easily prove that the majority vote works properly whenever $p<\frac{1}{2}$.

In Quantum Mechanics, errors are not so simple. Firstly, we cannot simply copy the quantum bits, because of the non-cloning theorem. Moreover, even if we could copy qubits, we would not be able to detect errors by simply measuring them. These limitations are enough to show us that our techniques must me changed if we wish to handle the quantum-mechanical world. But, if we wish to go a little further, we must consider that different kinds of errors may occur in this case. 

A qubit can suffer a phase flip, when $\alpha | 0 \rangle + \beta | 1 \rangle$ is changed to $\alpha | 0 \rangle - \beta | 1 \rangle$. Besides, we must remember that a qubit is continuous, so it can suffer a small error. Imagine a Bloch sphere where the state is accidentally rotated by a small angle. This error would not be detected if we consider only bit flips and phase flips. How can we protect our quantum computers against all these threats? Peter Shor, who created the code to correct arbitrary errors on a single qubit, answered this question~\cite{Shor96}. Nowadays, this code is known as the \textit{Shor code}. In order to understand it, we will analyze two ``modules'': the three-qubit bit flip code and the three-qubit phase flip code. The Shor code is a combination of these two ``modules''.

Before the explanation of these modules, it should be important to have a brief digression on the third postulate of Quantum Mechanics, which concerns to measurements. A measurement is characterized by an observable, $M$, such that $M=M^\dagger=(M^T)^\star$. So, the spectral decomposition of $M$ is:

\begin{equation}
M=\sum_m{P_m}.
\end{equation}

According to this postulate, the possible outcomes that may occur in an experiment are the eigenvalues of $M$, i.e., $m$. If the state before the measurement is $|\psi_{bef}\rangle$, then the probability that a result $m$ occurs is

\begin{equation}
p(m)=\langle \psi_{bef} | P_m | \psi_{bef} \rangle
\end{equation}
and the state after the measurement is

\begin{equation}
|\psi_aft \rangle = \frac{M_m|\psi\rangle}{\sqrt{\langle \psi_{bef} | P_m | \psi_{bef}}}.
\end{equation}

Now, we shall return to the study of the Shor code.

\subsection{The three-qubit bit flip code}

At first, we must encode our qubits, as follows:

\begin{eqnarray}
| 0 \rangle \;\;\mbox{becomes}\;\; | 0_L \rangle \equiv | 000 \rangle\label{eq:enc1}\\
| 1 \rangle \;\;\mbox{becomes}\;\; | 1_L \rangle \equiv | 111 \rangle\label{eq:enc2},
\end{eqnarray}
so that the qubit $|\psi\rangle = \alpha |0 \rangle + \beta |1 \rangle$ is encoded as $\alpha | 000 \rangle + \beta | 111 \rangle$.

If we wish to check if some error occurred, we use a measurement called \textit{syndrome diagnosis}. Its result is called the \textit{error syndrome}. We use four projection operators, each one corresponding to a bit flip in a specific qubit.

\begin{eqnarray}
P_0 \equiv |000 \rangle \langle 000| + |111 \rangle \langle 111| & \mbox{no error occurred}\\
P_1 \equiv |100 \rangle \langle 100| + |011 \rangle \langle 011| & \mbox{error on 1st qubit}\\
P_2 \equiv |010 \rangle \langle 000| + |101 \rangle \langle 101| & \mbox{error on 2nd qubit}\\
P_3 \equiv |001 \rangle \langle 000| + |110 \rangle \langle 110| & \mbox{error on 3rd qubit.}
\end{eqnarray}

Note that, when we measure an state $| \psi \rangle$ using the operator $P_i$, ($0 < i \leq 3$), we obtain $1$ if the $i$-th qubit is corrupted, and $0$ otherwise. Similarly, if we use the operator $P_0$, we obtain $1$ if the three qubits are correct, and $0$ otherwise. After the measurement, the qubit is not modified! This is possible because the syndrome diagnosis does not give us any information about the qubit \textit{per si} (i.e., the values of $\alpha$ and $\beta$).

\subsection{The three-qubit phase flip code}

The three-qubit bit flip code protects our qubit against the simplest kind of error. However, as we mentioned before, quantum bits are subject to much more errors than simple flips. The second type of noise that we will analyze is the phase flip. This error may change a qubit $\alpha|0\rangle + \beta|1\rangle$ into  $\alpha|0\rangle - \beta|1\rangle$.

The idea behind the three-qubit phase code is very similar to that used in the three-qubit bit flip code. Instead of encoding the qubit according to equations~\ref{eq:enc1} and~\ref{eq:enc2}, we must use

\begin{eqnarray}
| 0 \rangle \;\;\mbox{becomes}\;\; | 0_L \rangle \equiv | +++ \rangle\;\label{eq:enc3}\\
| 1 \rangle \;\;\mbox{becomes}\;\; | 1_L \rangle \equiv | --- \rangle,\label{eq:enc4}
\end{eqnarray}
where $|+\rangle \equiv (|0\rangle + |1\rangle)/\sqrt{2}$ and $|-\rangle \equiv (|0\rangle - |1\rangle)/\sqrt{2}$.

The only difference is that, in this code, we use the basis $\{|+\rangle,|-\rangle\}$ to encode the qubit, instead of the basis$\{|0\rangle,|1\rangle\}$. Remember that we can use the Hadamard operator (eq.~\ref{eq:hadamard}) to change between one basis and the other.

To perform the syndrome measurement, we only need to use the correct operators for the new basis.

\begin{equation}
P_i \;\;\mbox{becomes}\;\; P_i^{'} \equiv (H \otimes H \otimes H)P_i(H \otimes H \otimes H).
\end{equation}

If the reader is interested in demonstrations for the subject exposed so far in this section, we recommend the references~\cite{Chuang00} and~\cite{Preskill}.

\subsubsection{The Shor code}

The Shor code is a \textit{concatenation} of the two codes explained before. In the first step, we encode the qubit using the phase flip code (eq.~\ref{eq:enc3} and eq.~\ref{eq:enc4}). Consequently, we will have three qubits\footnote{We used three qubits in our example, but we could have used more. The same comment is valid for the bit flip code. Three is the minimum necessary in both cases.} from this point on. In the second step, we encode each of the three resulting qubits using the bit flip code (eq.~\ref{eq:enc1} and eq.~\ref{eq:enc2}). Finally, we have nine qubits representing the initial qubit.

\begin{eqnarray}
|0\rangle \;\;\mbox{becomes}\;\; |0_L\rangle \equiv \frac{(|000\rangle + |111\rangle)(|000\rangle + |111\rangle)(|000\rangle + |111\rangle)}{2\sqrt{2}}\label{eq:enc5}\\
| 1 \rangle \;\;\mbox{becomes}\;\; | 1_L \rangle \equiv \frac{(|000\rangle - |111\rangle)(|000\rangle - |111\rangle)(|000\rangle - |111\rangle)}{2\sqrt{2}}.\label{eq:enc6}
\end{eqnarray}

As we can see, this code is very expensive, especially if we consider the present status of experimental quantum computation. The best quantum computer ever built had only seven qubits! However, the Shor code is extremely important because it protects the qubit against arbitrary errors, not only bit flips or phase flips. The main lesson of this section is that, although quantum states can present a \textit{continuum} of errors, they can be repaired by detecting and correcting only a discrete subset of errors.

\section{Conclusions}

This paper is an introduction to one of the most important topics in Computer Science. Whenever we wish to perform a computation, we must know how to store a certain initial value, and then manipulate it until we achieve a result, when we can finally measure the system and obtain an answer.

We began by showing how we can perform simple operations in physical system, with classical and quantum logic gates. Then, we showed how to keep it reliable, with error detection and correction codes. Naturally, this is a vast area of knowledge, which could not be completely described in one single article. However, our goal here is to introduce this subject in a simple but effective way, helping to students and researchers from different areas of knowledge.

The references mentioned throughout this paper can be used to improve the comprehension of the subject.

\section*{Acknowledgements}

The authors thank Prof. J.A. Helay\"el-Neto (CBPF) and Dr. J.L. Acebal (PUC-Minas) for reading the manuscripts, and for providing helpful discussions. We thank the Group of Quantum Computation at LNCC, in particular Drs. R.~Portugal and F.~Haas, for the courses and stimulating discussions. We also thank the brazilian institution CNPq and the PIBIC program, for the financial support.

\end{document}